\documentclass[amsmath,twocolumn,amssymb,nofootinbib,prd]{revtex4-1}
\pdfoutput=1

\usepackage{graphicx}
\usepackage{hyperref}
\usepackage[outdir=./]{epstopdf}
\usepackage[caption=false]{subfig}

\begin{document}

\title{Conserved quantities and dual turbulent cascades in Anti--de Sitter spacetime}

\author{Alex Buchel}
\email{abuchel@perimeterinstitute.ca}
\affiliation{Department of Applied Mathematics, University of Western
Ontario, London, Ontario N6A 5B7, Canada}
\affiliation{Perimeter Institute for Theoretical Physics, Waterloo, Ontario N2L 2Y5,
Canada}
\author{Stephen R. Green}
\email{sgreen@perimeterinstitute.ca}
\affiliation{Perimeter Institute for Theoretical Physics, Waterloo, Ontario N2L 2Y5,
Canada}
\author{Luis Lehner}
\email{llehner@perimeterinstitute.ca}
\affiliation{Perimeter Institute for Theoretical Physics, Waterloo, Ontario N2L 2Y5,
Canada}
\author{Steven L. Liebling}
\email{steve.liebling@liu.edu}
\affiliation{Department of Physics, Long Island University, Brookville, NY 11548, U.S.A}

\date{\today}

\begin{abstract}
  We consider the dynamics of a spherically symmetric massless scalar
  field coupled to general relativity in Anti--de Sitter spacetime in
  the small-amplitude limit.  Within the context of our previously
  developed two time framework (TTF) to study the leading
  self-gravitating effects, we demonstrate the existence of two new
  conserved quantities in addition to the known total energy $E$ of
  the modes: The particle number $N$ and Hamiltonian $H$ of our TTF
  system.  Simultaneous conservation of $E$ and $N$ implies that weakly
  turbulent processes undergo dual cascades (direct cascade of $E$ and
  inverse cascade of $N$ or vice versa).  This partially explains the
  observed dynamics of 2-mode initial data.  In addition, conservation
  of $E$ and $N$ limits the region of phase space that can be explored
  within the TTF approximation and in particular rules out equipartition
  of energy among the modes for general initial data.  Finally, we
  discuss possible effects of conservation of $N$ and $E$ on late time
  dynamics.
\end{abstract}

\maketitle

\section{Introduction}

We are interested in the question of stability of Anti-de Sitter (AdS)
spacetime in general relativity.  Do small perturbations generically
collapse to a black hole or do they propagate forever?  In contrast
to Minkowski spacetime, field perturbations of AdS are effectively
confined by the asymptotically AdS boundary condition and cannot
dissipate by dispersing to null infinity.  Thus, even for arbitrarily
small perturbations, self-interactions eventually play a dominant role
in the dynamics, and the question of collapse is much more difficult.

In order to make the problem more tractable, we restrict to
spherically symmetric perturbations.  A real, minimally 
coupled, free, massless scalar field (coupled to gravity) is studied
to provide a dynamical degree of freedom.  This system was first
treated numerically in AdS$_4$
in~\cite{Bizon:2011gg} (for the non-spherically symmetric
gravitational case see Ref.~\cite{Dias:2011ss}, and for the complex scalar
field see Ref.~\cite{Buchel:2012uh}).  The authors of Ref.~\cite{Bizon:2011gg}
considered initial data with a Gaussian profile, and they varied the
amplitude of the initial pulse.  For sufficiently large pulse
amplitude, they found that black hole collapse occurs promptly,
consistent with expectations in asymptotically flat spacetimes.  Moreover, 
collapse is no longer prompt once the amplitude is decreased below a critical
threshold~\cite{Choptuik:1992jv}.  Instead, the pulse propagates to
infinity, and then is reflected back by the AdS boundary.  Once it
reaches the origin, it has another opportunity to collapse to a black
hole.  This time, however, gravitational focusing has caused the pulse
to become more peaked, and the chance of collapse is increased.  As
the amplitude is decreased further, multiple bounces may be required
before collapse.  Reference~\cite{Bizon:2011gg} found that collapse
always occurred even for very small-amplitude initial data, albeit
after a very large number of bounces.  This behavior motivated the conjecture
that collapse to a black hole was unavoidable in AdS regardless
of the initial perturbation and its strength (except for single-mode initial data~\cite{Bizon:2011gg,Maliborski:2013jca}).

The observed behavior is a manifestation of a weakly turbulent cascade,
where energy is transferred to short distance scales through
gravitational focusing.  A free, non-gravitating, spherically
symmetric, real scalar field in AdS$_4$ is characterized
by\footnote{Adding a mass or changing the spacetime dimension does not
  significantly alter this statement.  The mode functions and
  frequencies change, but the frequencies remain linear in $j$.} a set
of normal modes $e_j(x)$ ($j=0,1,2,\ldots$), with frequencies
$\omega_j=2j+3$.  In mode-space, the cascade corresponds to a transfer
of energy to high-frequency modes.  Since the frequency spectrum is
commensurate, the nonlinear gravitational interactions are resonant,
which means energy is transferred between modes quite readily.

This picture was complicated, however, by subsequent numerical
simulations~\cite{Buchel:2013uba} that showed that if the Gaussian
pulse profile was broadened (within a certain range), there was a
critical amplitude, below which collapse was no longer seen.  It was
argued that there was a nonlinear field dispersion process that
competes with the gravitational focusing.  Thus, if dispersion
dominates, then collapse can potentially be averted.  The problem was
thus more complicated than originally thought.  Because of the
intrinsic limitations of finite, numerical simulations, it demands a
more comprehensive perturbative analysis.

In previous work~\cite{Balasubramanian:2014cja}, we analyzed the
leading resonant interactions in a two-timescale analysis.  We
introduced a slow time $\tau\equiv \epsilon^2 t$, and expanded the scalar
field as $\phi(t,x) = \epsilon \phi_{(1)}(t,\tau,x)+O(\epsilon^3)$
(similar expansions apply for the metric). The general solution to the
leading-order scalar field equation is
\begin{equation}\label{eq:phi1}
  \phi_{(1)}(t,\tau,x)=\sum_{j=0}^\infty\left(A_j(\tau)e^{-i\omega_jt}+\bar{A}_j(\tau)e^{i\omega_jt}\right)e_j(x),
\end{equation}
where the functions $A_j(\tau)$ are---at this point---undetermined.
Were we to not introduce the slow time, $A_j$ would be constant, and
the resonant self-gravity interactions would lead to secular growths
and a breakdown of perturbation theory at
$O(\epsilon^3)$~\cite{Bizon:2011gg}.  However, as we showed
in~\cite{Balasubramanian:2014cja}, the secular growths can be
eliminated if we choose the mode amplitudes to satisfy the ``two time
framework''~(TTF) equations
\begin{equation}\label{eq:ttf}
  -2i\omega_j\frac{dA_j}{d\tau}=\sum_{klm}\mathcal{S}^{(j)}_{klm}\bar{A}_kA_lA_m,
\end{equation}
where the $\mathcal{S}^{(j)}_{klm}$ are (real) numerical coefficients
arising from overlap integrals of mode functions.  (We computed the
coefficients up to $j=j_{\text{max}}=47$.)  The TTF equations capture
(a) the effect of $\phi$ on the metric $g_{ab}$, and (b) the
backreaction of $g_{ab}$ on $\phi$.  In other words, $\phi$ can only
self-interact via $g_{ab}$ as an intermediary, and for this reason the
TTF equations have cubic interactions\footnote{TTF is universal when
  Einstein gravity is replaced with Gauss-Bonnet gravity
  \cite{Buchel:2014dba}; however it does require intrinsic
  self-interactions of the scalar field(s) to be subleading compared
  to gravitationally induced self-interactions.}.

The specific form of the equations~\eqref{eq:ttf} follows from the
fact that the only resonances that are actually present in our system
are those such that
\begin{equation}\label{rescon}
  \omega_j+\omega_k=\omega_l+\omega_m.
\end{equation}
Indeed from the basic structure of the Einstein-scalar field equations
it is {\em a priori} possible to have resonances satisfying
$\omega_j\pm\omega_k=\pm\omega_l\pm\omega_m$ for arbitrary sign
combinations, and this would introduce additional terms on the right
hand side of~\eqref{eq:ttf} (e.g., $\bar{A}_k\bar{A}_lA_m$, etc.).
However we found~\cite{Balasubramanian:2014cja} that in practice the
only non-zero terms are those of~\eqref{eq:ttf} (see
also~\cite{Bizon:2011gg}).  This statement was rigorously proven
in~\cite{Craps:2014vaa}.

We argued in~\cite{Balasubramanian:2014cja} that solutions to the TTF
equations provide a good approximation to solutions of the full system
in the limit $\epsilon\to0$ and for time scales $t\sim1/\epsilon^2$.
By solving the system of coupled ODEs~\eqref{eq:ttf}, one can study,
e.g., the transfer of energy $E_j(\tau)=4\omega_j^2|A_j(\tau)|^2$
between the modes---without having to consider the rapid normal-mode
harmonic oscillations.  We checked previously that, while $E_j(\tau)$
can have very nontrivial time-dependence, the sum $E\equiv\sum_jE_j$
is conserved.

As we will describe in section~\ref{sec:conservedquantities}, there are two additional quantities that
are conserved by the system of TTF equations\footnote{These properties
  apply also for massive fields and in different dimensions, since
  the structure of the TTF equations is the same.}.  The first,
\begin{equation}
  N\equiv\sum_j4\omega_j|A_j|^2,
\end{equation}
can be interpreted as the ``particle number'' of the field.  In terms
of spacetime fields, this is the charge current of the positive
frequency part of $\phi$.  (Since $\phi$ is real, the charge current
of the field itself vanishes.)  The second additional conserved
quantity is the Hamiltonian function $H$ of our TTF
system~\eqref{eq:ttf}.  This quantity is quartic in $A_j$ and appears
to represent the interaction energy of the TTF fields. All three
quantities ($N$, $E$ and $H$) are only well-defined and conserved at
the perturbative level captured by TTF, and it is not clear to what
degree they extend to the fully nonlinear theory.


In section~\ref{sec:dualcascade}, we describe how the presence of
multiple conserved quantities (as opposed to solely the energy)
implies that the turbulent dynamics must be more complex than
originally thought~\cite{Bizon:2011gg}. This is well-known, for
example, in the context of incompressible, inviscid two-dimensional hydrodynamics, where the
existence of a second conserved quantity induces an inverse
cascade of energy, in contrast to the standard direct energy cascade
in higher dimensions~\cite{Kraichnan:1967,Boffetta:2012}.  Likewise,
in the TTF system~\eqref{eq:ttf}, conservation of $E$ and $N$ implies
that the energy cannot all be transferred to high-$j$ modes, as this
would violate conservation of $N$.  As higher frequency modes have
higher energy per particle, a given amount of energy gives rise to a
lower population of particles.  To conserve particle number as energy
transfers to high modes, lower modes must also become more
populated---an effect we have previously observed for 2-mode initial
data~\cite{Balasubramanian:2014cja}.  Thus, if there is a direct
cascade of $E$, there must simultaneously be an inverse cascade of
$N$ and vice versa.  We therefore have an alternative explanation for
the process of competing dispersion and focusing described
in~\cite{Buchel:2013uba}.

Finally, in section~\ref{sec:noneq} we explore the implications of
conservation laws on possible end-states of evolution.  Simultaneous
conservation of $N$ and $E$ excludes certain regions of phase space
and,  in particular, generically, does not allow for energy
equipartition.  Instead, one can use $N$ and $E$ conservation to
associate our initial data to recently uncovered quasi-periodic
solutions~\cite{Balasubramanian:2014cja}.  In particular, if a given
solution lies sufficiently close to its associated quasi-periodic
solution then it is really a perturbation about that solution.  It
follows that if the solution lies within the radius of stability of
the quasi-periodic solution, then it cannot collapse.  This is a
generalization of the claim that broad Gaussian initial data is
``close'' to stable time-periodic solutions~\cite{Maliborski:2013ula}.
An alternative possibility is that a non-collapsing behavior can be
described by stochasticity among the stable basins of these
quasi-periodic solutions~\cite{chirikov2} as argued by Ref.~\cite{IC} 
for the Fermi-Pasta-Ulam problem~\cite{Fermi:1955:SNP}.

In this paper we follow the notation and definitions
of~\cite{Balasubramanian:2014cja}. We note that as we were preparing
this manuscript a separate work also identified the new conserved
quantities~\cite{Craps:2014jwa} within the perturbative resummation
technique presented in~\cite{Craps:2014vaa} (see
also~\cite{Basu:2014sia} for the case of a non-gravitating probe
scalar). As discussed in~\cite{Craps:2014vaa}, the equations of motion
obtained to third order are identical to those from the TTF approach.

\section{Conserved quantities}\label{sec:conservedquantities}

\subsection{Total mode energy and total particle number}

By using the TTF equations~\eqref{eq:ttf} and the specific
coefficients $\mathcal{S}^{(j)}_{klm}$ we derived up to
$j=j_{\text{max}}=47$, it is easy to check directly that
$E=\sum_j4\omega_j^2|A_j|^2$ and $N=\sum_j4\omega_j|A_j|^2$ are
conserved in time (up to $j=j_{\text{max}}$).  (We did this for $E$
in~\cite{Balasubramanian:2014cja}, and this was also how we first
identified $N$.)  However, the conservation laws are more general in
that they follow simply from symmetry properties of
$\mathcal{S}^{(j)}_{klm}$ (and not the specific values these
coefficients), as we will show below.

Because of its similarity to expressions in QFT, we interpret the
quantity $N$ as the total ``particle number.''  However, we note that
this is a continuous system, and there are no discrete particles.


\subsubsection{Derivation}

To proceed, we assume only certain basic symmetry properties, which
hold\footnote{For $j,k,l,m$ pairwise distinct (e.g.,
  $j,k,l,m=0,1,0,1$) symmetries (i) and (iii) only hold in the gauge
  where the metric function $\delta=0$ at the AdS boundary.  This
  gauge was used in~\cite{Buchel:2012uh,Craps:2014jwa},
  whereas \cite{Bizon:2011gg,Balasubramanian:2014cja,Craps:2014vaa} set $\delta=0$
  at the origin. (For 1, 3 or 4 distinct indices $S^{(j)}_{klm}$
  satisfies all the symmetries above in both gauges).  Since the
  symmetries are needed in Sec.~\ref{sec:hamiltonian} to write down a
  Hamiltonian formulation, we therefore assume we are working in the
  gauge where $\delta=0$ at the AdS boundary.  See
  \cite{Craps:2014jwa} for further discussion of this point.} for the
coefficients $\mathcal{S}^{(j)}_{klm}$,
\renewcommand{\labelenumi}{(\roman{enumi})}
\begin{enumerate}
\item $\mathcal{S}^{(j)}_{klm}=\mathcal{S}^{(k)}_{jlm}$,
\item $\mathcal{S}^{(j)}_{klm}=\mathcal{S}^{(j)}_{kml}$,
\item $\mathcal{S}^{(j)}_{klm}=\mathcal{S}^{(l)}_{mjk}$,
\item $\mathcal{S}^{(j)}_{klm}=0$ unless $j+k=l+m$.
\end{enumerate}
\renewcommand{\labelenumi}{\underline{Case \theenumi.}}  Given
symmetries (i)--(iii), we now show that conservation of $E$ is
precisely equivalent to (iv).  The latter reflects the resonance
condition \eqref{rescon}.  Conservation of $N$ follows automatically
from (i)--(iii) alone.

We first note that, as a result of~\eqref{eq:ttf},
\begin{eqnarray}
  \frac{d}{d\tau}|A_j|^2&=&\bar{A}_j\frac{dA_j}{d\tau}+A_j\frac{d\bar{A}_j}{d\tau}\nonumber\\
  &=&\frac{i}{2\omega_j}\sum_{klm}\mathcal{S}^{(j)}_{klm}\left(\bar{A}_j\bar{A}_kA_lA_m-A_jA_k\bar{A}_l\bar{A}_m\right)\nonumber\\
  &=&\frac{1}{\omega_j}\sum_{klm}\mathcal{S}^{(j)}_{klm}\Im \left(A_jA_k\bar{A}_l\bar{A}_m\right).
\end{eqnarray}
Conservation of $E$ and $N$ are then
\begin{eqnarray}
  \label{eq:dEdt}0&=&\frac{dE}{d\tau}=\sum_{jklm}4\omega_j\mathcal{S}^{(j)}_{klm}\Im \left(A_jA_k\bar{A}_l\bar{A}_m\right),\\
  \label{eq:dNdt}0&=&\frac{dN}{d\tau}=\sum_{jklm}4\mathcal{S}^{(j)}_{klm}\Im \left(A_jA_k\bar{A}_l\bar{A}_m\right),
\end{eqnarray}
for all possible values of the $A_j$.

Consider the energy $E$ first.  We look at the contribution of terms
with particular choices of indices.
\begin{enumerate}
\item $j,k,l,m$ all identical

  We have $\Im
  \left(A_jA_k\bar{A}_l\bar{A}_m\right)=\Im\left(|A_{j}|^4\right)=0$,
  so this type of term does not contribute, and the right hand side of
  \eqref{eq:dEdt} vanishes automatically, imposing no constraints on
  the coefficients $\mathcal{S}_{klm}^{(j)}$.

\item $j=l=m$, $k$ distinct

  Imposing the symmetries (i)--(iii), terms with these particular
  indices contribute
  \begin{eqnarray}
    \frac{dE}{d\tau}&\supset& 8\omega_jS^{(j)}_{jjk}\Im(A_j^2\bar{A}_j\bar{A}_k)\nonumber\\ &&+4(\omega_j+\omega_k)S^{(j)}_{kjj}\Im(A_jA_k\bar{A}_j^2)\nonumber\\
    &=&4\Im(A_j^2\bar{A}_j\bar{A}_k)\left[2\omega_jS^{(j)}_{jjk}-(\omega_j+\omega_k)S^{(j)}_{kjj}\right].
  \end{eqnarray}
  This vanishes if (iv) holds.

  Conversely, assume that $E$ is conserved.  The right hand side of
  \eqref{eq:dEdt} must vanish for all values of the $A_j$'s.  Now
  setting only $A_j$, $A_k$ to be nonzero implies
  $S^{(j)}_{jjk}=S^{(j)}_{kjj}=0$.

\item $j=l$, $k=m$ distinct

  This contributes
  \begin{equation}
    \frac{dE}{d\tau}\supset4\Im(A_j^2\bar{A}_k^2)\left[\omega_j\mathcal{S}^{(j)}_{jkk}-\omega_k\mathcal{S}^{(k)}_{kjj}\right].
  \end{equation}
  Since $\omega_j\ne\omega_k$, (iv) here is equivalent to the vanishing of this contribution, as above.

\item $j=m$, $k$, $l$ distinct

  The symmetries imply that the contribution to $dE/d\tau$ is
  \begin{equation}
    \frac{dE}{d\tau}\supset4\Im(A_j^2\bar{A}_k\bar{A_l})\mathcal{S}^{(j)}_{jkl}\left[2\omega_j-(\omega_k+\omega_l)\right].
  \end{equation}
  Vanishing of the multiplier implies either
  $\mathcal{S}^{(j)}_{jkl}=0$ or $2\omega_j=\omega_k+\omega_l$.

\item $j,k,l,m$ all distinct

  In this last case, the contribution is
  \begin{eqnarray}
    \frac{dE}{d\tau}&\supset&8\left\{\Im(A_jA_k\bar{A}_l\bar{A}_m)\mathcal{S}^{(j)}_{klm}\left(\omega_j+\omega_k-\omega_m-\omega_l\right) \right.\nonumber\\
    &&+\Im(A_jA_l\bar{A}_k\bar{A}_m)\mathcal{S}^{(j)}_{lkm}\left(\omega_j+\omega_l-\omega_k-\omega_m\right)\nonumber\\
    &&\left.+\Im(A_jA_m\bar{A}_k\bar{A}_l)\mathcal{S}^{(j)}_{mkl}\left(\omega_j+\omega_m-\omega_k-\omega_l\right)\right\}.
  \end{eqnarray}
  Since the coefficients containing the $A_j$'s can be made to be
  nonzero independently, we again see the equivalence of symmetry (iv)
  and the vanishing of the term.
\end{enumerate}

Thus, it follows that conservation of $E$ is precisely equivalent to
condition (iv), given (i)--(iii).  Conservation of $N$ can be analyzed by
replacing all $\omega_j$'s in the expressions above by 1.  All terms
then vanish by (i)--(iii).$\square$

\subsubsection{Discussion}

From a spacetime perspective, conservation of $N$ and $E$ is somewhat
surprising.  Indeed, $E$ is {\em not} the total energy of the system,
as it neglects interaction energy.  Meanwhile, our scalar field is
real so we do not expect any sort of associated charge $N$.  What this
TTF analysis shows is that at a low (but nonlinear) perturbative
order, and for times $t\sim1/\epsilon^2$, these quantities are in fact conserved.  

To see how $N$ and $E$ arise as spacetime quantities, consider a treatment of
$\phi$ at the {\em linear} level.  It should be kept in mind that at
the linear level, there is no exchange of energy between modes, so
conservation of $N$ and $E$ is trivial.  Nevertheless, this treatment
is useful as there is a precise correspondence between our definitions
of $E$ and $N$ and spatial integrals of field quantities.  Indeed, at this
order, $E$ is just the spatial integral of the time-time
component of the stress-energy of $\phi^{(1)}$, living in an exact AdS
background,
\begin{equation}\label{eq:Eint}
  E =\int_0^{\pi/2} \left(|\partial_t\phi^{(1)}|^2+|\partial_x\phi^{(1)}|^2\right)\tan^2x\,\mathrm{d}x.
\end{equation}
Plugging in the expression~\eqref{eq:phi1} with each
$A_j(\tau)=\text{constant}$, we recover the standard mode sum
expression for $E$.

For the particle number, split the field into positive and negative
frequency parts, $\phi^{(1)}=\phi^{(1)}_{+}+\phi^{(1)}_{-}$, with
\begin{equation}
  \phi^{(1)}_{+}=\sum_j A_j e^{-i\omega_j t}e_j(x).
\end{equation}
The positive frequency part of $\phi^{(1)}$ {\em is} a complex field,
so one can define its charge current.  One may verify that the
integral of its time component is the particle number,
\begin{equation}
  N=2i\int_0^{\pi/2} \left(\overline{\phi^{(1)}_+}\partial_t\phi^{(1)}_+ - \phi^{(1)} \partial_t \overline{\phi^{(1)}_+}\right) \tan^2 x \,\mathrm{d}x.
\end{equation}

Beyond linear order it is not obvious how to define the positive
frequency part of the field, nor is it obvious that $N$ and $E$ are
conserved from a spacetime perspective.  Fortunately, TTF provides a
way to perturbatively define quantities that are conserved.

\subsection{Hamiltonian}\label{sec:hamiltonian}




In addition, the TTF system is Hamiltonian.  To see this, we first
rescale our fields by defining $\hat{A}_j=\sqrt{\omega_j}A_j$.  Then
the Hamiltonian takes the form
\begin{equation}\label{eq:hamiltoniandef}
  H=-\frac{1}{4}\sum_{jklm}\frac{\mathcal{S}^{(j)}_{klm}}{\sqrt{\omega_j\omega_k\omega_l\omega_m}}\bar{\hat{A}}_j\bar{\hat{A}}_k\hat{A}_l\hat{A}_m.
\end{equation}
It may be verified that Hamilton's equations
\begin{equation}
  i\frac{d\hat{A}_i}{d\tau} = \frac{\partial H}{\partial \bar{\hat{A}}_i}
\end{equation}
reproduce the equations of motion~\eqref{eq:ttf}.  Indeed, Hamilton's
equations give
\begin{eqnarray}
  &&-2i\omega_i\frac{dA_i}{d\tau}\nonumber\\
  &=&-2i\sqrt{\omega_i}\frac{d\hat{A}_i}{d\tau}\nonumber\\
  &=&-2i\sqrt{\omega_i}\frac{\partial H}{\partial \bar{\hat{A}}_i} \nonumber\\
  &=& \frac{\sqrt{\omega_i}}{2} \sum_{jklm} \frac{\mathcal{S}^{(j)}_{klm}}{\sqrt{\omega_j\omega_k\omega_l\omega_m}}\left(\delta^i_j\bar{\hat{A}}_k\hat{A}_l\hat{A}_m+ \bar{\hat{A}}_j\delta^i_k\hat{A}_l\hat{A}_m\right)\nonumber\\
  &=&\sum_{jkl}\frac{\mathcal{S}^{(i)}_{jkl}}{\sqrt{\omega_j\omega_k\omega_l}}\bar{\hat{A}}_j\hat{A}_k\hat{A}_l\nonumber\\
  &=&\sum_{jkl}\mathcal{S}^{(i)}_{jkl}\bar{A}_jA_kA_l.
\end{eqnarray}
[On the second last line we used the symmetry property (i) of
$\mathcal{S}^{(j)}_{klm}$.]  $H$ itself is thus a conserved quantity.

It may be tempting to try to associate the Hamiltonian with the
next-order contribution to the ADM mass $M$ of the spacetime.  (Note
that $H$ itself is not the total energy because we have performed a
two-time expansion to factor out the fast time.)  Indeed, the ADM mass
is a conserved quantity in general relativity, and it includes all of
the kinetic and potential energy in the field $\phi$, as well as the
metric.  The energy $E$, by contrast, only contains the energy of
$\phi$ to quadratic order, while the expression
\eqref{eq:hamiltoniandef} for $H$ certainly looks like a potential
energy that could be the fourth order contribution to $M$.  However,
one would also expect additional contributions to $M$ at quartic
order, such as contributions from higher order TTF corrections
(involving additional slower time variables).  Further analysis would
therefore be required to elucidate any relationship between $H$ and
the ADM mass.

It is somewhat surprising that $E$ and $H$ are conserved
independently.  This could simply be an unphysical artifact of our
expansion, or it could be indicative of a whole family of higher order
conserved quantities, which would indicate integrability of the
system.  It may be illuminating to numerically monitor the behavior of
$E$ and $H$ within the fully nonlinear system, at least over time
scales where we expect validity of TTF ($t\sim1/\epsilon^2$).

\section{Dual cascade}\label{sec:dualcascade}

For the remainder of this paper, we restrict our analysis to examining the implications of simultaneous
conservation of $E$ and $N$.  Within the context of turbulence, the
presence of a second conserved quantity (in addition to the energy)
indicates the occurrence of dual cascades.  That is, if one
quantity is cascading to higher modes, the other must be
simultaneously cascading to low modes.  As noted in the introduction, this has been observed, for
instance, in inviscid incompressible two-dimensional fluid dynamics\footnote{This cascading
behavior is also present in the viscous/compressible case,
but its analysis is sightly more involved.}, where the {\em enstrophy} $\Omega$
(the integral of the vorticity squared) is conserved in addition to the energy ${\cal E}$.  
It can be shown that enstrophy undergoes a direct cascade, which forces energy to cascade
in the opposite direction, leading to the formation of increasingly large
vortices~\cite{Kraichnan:1967,Boffetta:2012}.

In momentum space,
\begin{equation}
{\cal E} = \int E(k)\,\mathrm{d}k, ~~~~ ~~~~ \Omega = \int k^2 E(k) \,\mathrm{d}k,
\end{equation}
with $E(k)$ the fluid energy at wavenumber $k$.  On the other hand, for the TTF system we have
\begin{equation}
E=\sum_j E_j, ~~~~~~~~~~ N=\sum_j(2j+3)^{-1}E_j.
\end{equation}
There is thus a natural parallel in both cases. However, the
difference in the exponent of the wave number ($k$ or $j$) has an important
consequence. Namely, in the hydrodynamic case, energy must flow to
longer wavelengths and enstrophy to higher ones irrespective of the
wavenumber (assuming there is no upper limit to the wavelengths
allowed). On the other hand, in the scalar case energy flows both to
low and high wavenumbers by similar amounts in the high wavenumber
regime (where the 3 can be neglected compared to $2j$).  

For now, let us concentrate on understanding the dual cascade process in the
scalar case.  Suppose that, under time
evolution, all of the energy $E$ could be transferred to higher-$j$ modes.
In that case, since higher-$j$ modes have more energy per particle
($E_j/N_j = \omega_j$), this process would violate particle number
conservation.  Instead, in order to conserve $N$
some energy must
transfer to lower-$j$ modes, which are less energetic.

\begin{figure}[t]
\begin{center}
 \includegraphics[width=3.3in,clip]{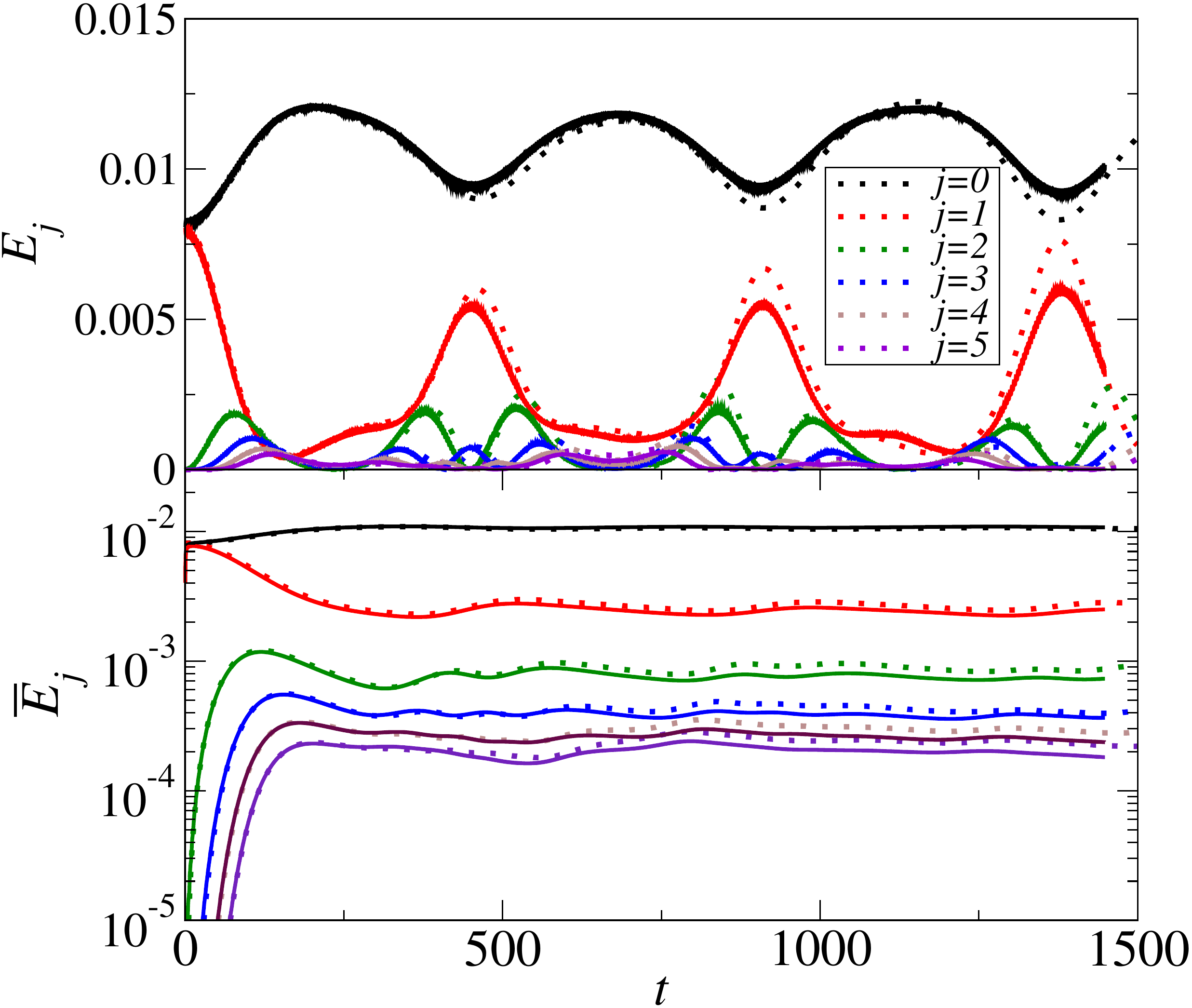}
\end{center}
\caption{Full numerical (solid) and TTF (dotted) energy (top panel)
  and running time-average energy (bottom panel) per mode, for 2-mode
  equal-energy initial data.  The transfer of energy from mode $j=1$
  into mode $j=0$ is a consequence of $N$ conservation. Note that
  there is some uncertainty as to whether collapse occurs around
  $t\approx 1080$~\cite{Bizon:2014bya}, but this is not relevant to
  our discussion of dual turbulent cascades.}
 \label{fig:2mode-energy}
\end{figure}
This phenomenon has in fact already been observed in our previous
work~\cite{Balasubramanian:2014cja}, when we studied the evolution of
initial data with the energy distributed evenly between the two lowest
modes.  We reproduce in figure~\ref{fig:2mode-energy} our prior
results.  Notice that the energy initially flows {\em into} mode $j=0$
from mode $j=1$.  This behavior is now understood in light of the
second conserved quantity, for if all the energy were to flow to
higher modes, then $N$ would not be conserved.  The correspondence
between full numerics and TTF shown in figure~\ref{fig:2mode-energy}
also indicates that conservation of $N$ is not merely an effect of our perturbation
scheme.

\section{Non-equipartition of energy and stability}\label{sec:noneq}

That the Fermi-Pasta-Ulam (FPU) system of nonlinearly coupled
oscillators did not reach equipartition for small initial energy was
quite surprising and was seminal in the development of nonlinear
dynamics~\cite{Fermi:1955:SNP,2005Chaos..15a5104B,fpubook}. In the
current context of a scalar field in AdS, if a black hole does not
form, then the question of thermalization of the holographically dual
CFT is complicated. Recall that evaporation of small black holes gives
rise to thermal states and so naturally bridges the initial state on
the dual CFT to its final thermal one, represented by a partially
evaporated black hole in equilibrium with surrounding Hawking
radiation~\cite{Horowitz:1999uv}.  Consequently, failure to yield a
black hole raises the question of whether and how the corresponding
CFT will achieve a thermal state. One should take note here that our
calculations within Einstein gravity are limited to the classical
regime (thus ignoring $1/N^2$ corrections) and also ignore possible
higher curvature modifications\footnote{Within the class of
  Gauss-Bonnet corrections, it has been recently pointed out that many
  configurations avoid black hole
  collapse~\cite{Deppe:2014oua,Buchel:2014dba}.}.

With such caveats in mind, one can show that within TTF, the fact that
both $N$ and $E$ are conserved implies that energy equipartition
cannot occur. To see this, assume some initial state has energy $E$
and particle number $N$.  Then, if equipartition were to occur, each
mode would have energy $E_j=E/(j_{\text{max}}+1)$ (where we truncate
our system at $j=j_{\text{max}}$).  But that would imply that the total particle number is
\begin{eqnarray}
  N_{\text{final}} &=& \sum_{j=0}^{j_{\text{max}}} \frac{E_j}{\omega_j} = \sum_{j=0}^{j_{\text{max}}}\frac{E}{\omega_j(j_{\text{max}}+1)}\nonumber\\
  &=& \frac{H_{j_{\text{max}}+\frac{3}{2}}-2+\log4}{2 (j_{\text{max}}+1)}E,
\end{eqnarray}
where $H_n$ is the $n$th harmonic number.  Unless  finely tuned initially, this value will not equal $N$ and therefore will be excluded dynamically.

\begin{figure}[t]
\centering
  \begin{tabular}{c}
    \subfloat[$\lambda=0$]{\label{fig:qp-2modeA}\includegraphics[clip]{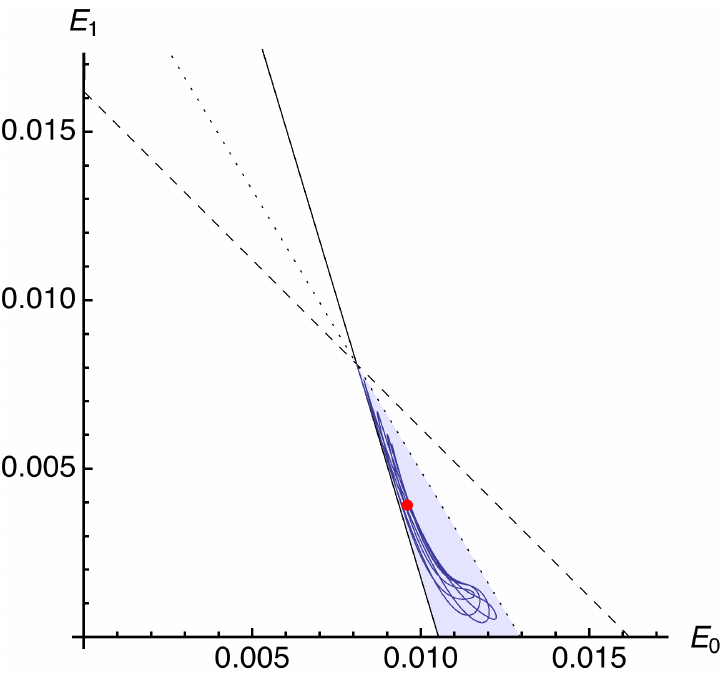}}\\
    \subfloat[$\lambda=0.5$]{\label{fig:qp-2modeB}\includegraphics[clip]{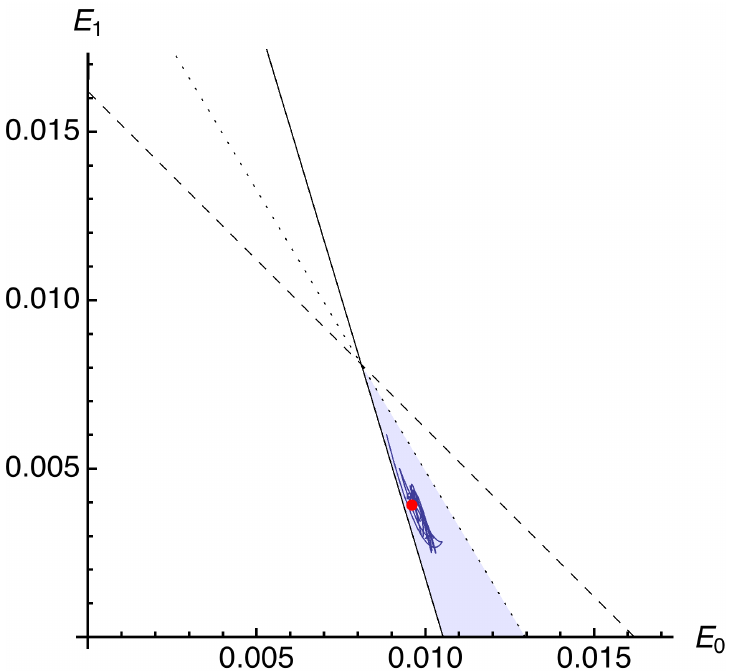}}
  \end{tabular}
\caption{Paths of $E_0$ and $E_1$ for (a) 2-mode initial data of figure~\ref{fig:2mode-energy} (blue curve; TTF solution with $j_{\text{max}}=47$) and (b) interpolation between 2-mode and quasi-periodic initial data.  Red dot represents the quasi-periodic solution with the same values of total $E$ and $N$ as the 2-mode solution. Solution is constrained by $E$ and $N$ conservation to lie within the blue shaded region.}
 \label{fig:qp-2mode}
\end{figure}
In fact, one can go further and place bounds on regions of phase space
that are excluded by conservation of $N$ and $E$.  We illustrate this
in figure~\ref{fig:qp-2mode}.  The blue curve in
figure~\ref{fig:qp-2modeA} represents the path of the 2-mode solution
of figure~\ref{fig:2mode-energy} in $(E_0,E_1)$-space.  It is confined
to lie within the shaded blue region between the lines drawn on
the plot.  The two upper boundary lines arise because of $E$ (dashed line) and $N$ (dotted line)
conservation,
\begin{eqnarray}
  E_0+E_1&\le& E,\\
  \frac{E_0}{3}+\frac{E_1}{5}&\le&N=\frac{4}{15}E.
\end{eqnarray}
The lower bound (solid line) arises because only a certain amount of energy can
flow to the higher modes.  The energy that can flow to higher modes is
maximized if it flows into mode $j=2$ because its particles are least
energetic.  So we write $E$ and $N$ conservation as
\begin{eqnarray}
  E_0+E_1+E_2+\sum_{j>2}E_j&=&E,\\
  \frac{E_0}{3}+\frac{E_1}{5}+\frac{E_2}{7}+\sum_{j>2}\frac{E_j}{2j+3}&=&N=\frac{4}{15}E.
\end{eqnarray}
Eliminating $E_2$, we obtain our bound,
\begin{eqnarray}
  \frac{4}{3}E_0+\frac{2}{5}E_1&=&\frac{13}{15}E+\sum_{j>2}\left(1-\frac{7}{2j+3}\right)E_j\nonumber\\
  &\ge&\frac{13}{15}E,
\end{eqnarray}
since the last term on the first line is non-negative.

Given the bounds, it is clear that energy equipartition is not
possible, since equipartition occurs at
$(E_0,E_1)=(1/(j_{\text{max}}+1),1/(j_{\text{max}}+1))\to (0,0)$ as
$j_{\text{max}}\to\infty$.  More generally, the bounds arising from
the conservation laws limit the amount of energy that can cascade to
high modes.

As equipartition is ruled out, one can consider other possible late
time configurations where some energy has transferred to high-$j$
modes.  For example, power laws of the form $E_j\sim (j+1)^{-\alpha}$
were observed in numerical evolutions just prior to collapse for
Gaussian initial data~\cite{Maliborski:2013via,deOliveira:2012dt}.  It is
straightforward to determine, for a given $\alpha$, the corresponding
value of $E/N$ for the configuration.  Larger $\alpha$ (steeper power
laws) correspond to smaller values of $E/N$, with $E/N\searrow3$ as
$\alpha\to\infty$ in AdS${}_4$. It is intriguing to note that the
$\sigma=1/16$ Gaussian initial data of~\cite{Bizon:2011gg} is
characterized by the power law with $\alpha=1.15$; not that far from
the value $\alpha=1.2$ observed just prior to
collapse~\cite{Maliborski:2013via}.  The $\sigma=0.4$ Gaussian initial
data of~\cite{Buchel:2013uba}---that appears to avoid collapse---has
$E/N=3.43$, which corresponds to a much steeper power law of
$\alpha=2.60$.  The 2-mode equal-energy configuration has $E/N=3.75$,
corresponding to $\alpha=2.15$.  The quantity $E/N$ may therefore
indicate whether given initial data is likely to collapse for small
initial amplitude.  However this cannot be the full story, since
single-mode initial data has $E/N=\omega_j$ (which can be arbitrarily
large for large $j$), and is expected not to
collapse~\cite{Maliborski:2013jca}.


In light of the fact that we know that both $N$ and $E$ are conserved,
we turn now to a re-examination of a class of quasi-periodic
solutions we previously identified within the context of TTF.
In~\cite{Balasubramanian:2014cja}, we looked for solutions with
harmonic $\tau$-dependence in each mode, $A_j(\tau) = \alpha_j
e^{-i\beta_j \tau}$, with $\beta_j\in\mathbb{R}$ (so that the energy
in each mode is constant).  We found that, given a particular discrete choice of a
dominant mode, there exists a two-parameter family of quasi-periodic
solutions that generalize\footnote{The periodic solutions
  of~\cite{Bizon:2011gg} were extended to higher nonlinear order
  in~\cite{Maliborski:2013jca}.  We have not attempted this analysis
  for our quasi-periodic solutions, however numerical evidence
  indicates stability in the full theory.}  the one-parameter periodic
solutions of~\cite{Bizon:2011gg}.  These solutions have approximately
exponential energy spectrum to both sides of the dominant mode (see
figure 1 of~\cite{Balasubramanian:2014cja}), with the exponential
decay rate one of the two parameters (the other being the overall
amplitude).  We could only find such solutions for sufficiently fast
exponential fall-off.  Under fully nonlinear numerical evolution, the
solutions appeared to be stable.

Instead of taking the two parameters characterizing a quasi-periodic
solution to be the amplitude and the exponential decay rate of the
energy spectrum, we can equivalently take them to be $N$ and $E$.
Thus, given a choice of a dominant mode, and values for $N$ and $E$
within a suitable range, there is a quasi-periodic TTF solution that
appears to be stable under the full numerical evolution.

For our 2-mode initial data, we can extract $N$ and $E$, and
construct a quasi-periodic solution with these values, based around
the dominant mode $j=0$.  This is presented as the red dot in
figure~\ref{fig:qp-2modeA}. We see that the 2-mode solution seems to
oscillate around the quasi-periodic solution rather than filling the
entire shaded blue region available to it.  Following the $q$-breather
approach to understanding the FPU
problem~\cite{Flach:2005,Flach:2006}, we can ask whether the 2-mode
solution should be thought of as a perturbation about the associated
quasi-periodic solution (which itself is a generalization of a
$q$-breather to the case with both $N$ and $E$ conserved).  To study
this possibility, we considered initial data that interpolates (with
parameter $\lambda$) between the 2-mode initial data ($\lambda=0$)
and the quasi-periodic initial data ($\lambda=1$).  In
figure~\ref{fig:qp-2modeB} we present the results for $\lambda=0.5$.
Notice that the solution is confined to a smaller region in
$(E_0,E_1)$-space around the quasi-periodic solution.  This is to be
expected if it can be treated as a perturbation about the
quasi-periodic solution.

As $\lambda$ is varied from 0 to 1, the dynamics of $E_j(\tau)$
smoothly deform to smaller oscillations, but maintain the overall
structure of recurrences seen in figure~\ref{fig:2mode-energy}.  In
particular, the time of the first recurrence changes by less than a
factor of two.  This supports the claim that the 2-mode solution can
be regarded as a perturbation (albeit a large one) of the quasi-periodic
solution.  In that case, the recurrences are merely oscillations about the
quasi-periodic solution. (This doesn't preclude collapse, as the 2-mode solution can in principle be an unstable perturbation of the quasi-periodic solution\footnote{There
  is some disagreement in the literature as to whether our particular 2-mode data collapses or not in numerical studies~\cite{Balasubramanian:2014cja,Bizon:2014bya}.}.) If in the full theory, the quasi-periodic
solutions persist and are stable, and if a given solution lies within
the radius of stability, then it must avoid collapse. 

Alternatively, as argued in~\cite{IC,chirikov2}, this
behavior could be illustrating stochasticity between the stability
regions of each individual mode. Both of these options are plausible
explanations for the possible stability of 2-mode data.

\section{Final comments}

The question of stability of AdS has recently received a great deal of
attention, as it is of interest within general relativity, and it has
holographic applications.  While the full answer to this question is
still outstanding, a combination of numerical simulations and
perturbative studies has provided many important insights (see
e.g.~\cite{Bizon:2011gg,Dias:2011ss,Buchel:2012uh,Dias:2012tq,Buchel:2013uba,Maliborski:2013jca,Balasubramanian:2014cja,Buchel:2014dba,Bizon:2014bya,Craps:2014vaa,Dimitrakopoulos:2014ada,Craps:2014jwa}).
Using the perturbative approach introduced
in~\cite{Balasubramanian:2014cja}, we studied conserved quantities in
the spherically symmetric Einstein-scalar system.  We identified the
total mode-energy $E$, the Hamiltonian $H$, and the particle number
$N$ as conserved quantities, and we explored their dynamical effects.
Based on our understanding of TTF, we expect these quantities to be
conserved for timescales $t\sim1/\epsilon^2$ in the full theory, and
so they are particularly useful for studying small-amplitude
perturbations. \

We have shown that conservation of $E$ and $N$ together imply that an
inverse cascade of particles must accompany any energy flow to higher
wavenumbers and vice versa.  Conservation of $E$ and $N$ also causes
parts of phase space (including energy equipartition) to be dynamically
excluded. Additionally, we have discussed the possibility of
quasi-periodic $q$-breathers as providing yet further examples of
stability islands.

Conservation laws therefore play a very important role in constraining
the dynamics of the system.  With three constants of motion already
identified for the problem of spherical collapse in AdS, it is
tantalizing to imagine that others are still waiting to be discovered.
Numerical evidence suggests there are several families of initial data
that do not lead to black hole formation.  Non-thermalization in
1-dimensional systems such as FPU chains is evidence that the system
may be integrable; i.e., that it contains an infinite number of
conserved quantities.

Integrable systems in more than 1 spatial dimension, however, are
quite rare, and so in that sense instability is more likely for the full
gravitational problem in the absence of spherical symmetry.  This is
related to the fact that the density of states grows more rapidly with
energy in higher dimensions.  On the other hand, black hole formation
is generally thought to be easiest {\em with} spherical symmetry.
This tension remains to be resolved.

\acknowledgments

We would like to thank D.~Abanin, J.~Santos, P.~Vieira,
J.~R.~Westernacher-Schneider, and H.~Yang for interesting discussions.
During the course of our work, Ref.~\cite{Craps:2014jwa} was posted,
which independently identified the two new conserved quantities, and
presented a thorough analysis. This work was supported by the NSF
under grant PHY-1308621~(LIU), by NASA under grant NNX13AH01G, by
NSERC through a Discovery Grant (to A.B. and L.L.) and by CIFAR (to
L.L.).  Research at Perimeter Institute is supported through Industry
Canada and by the Province of Ontario through the Ministry of Research
\& Innovation.

\bibliography{references.bib}

\end{document}